\documentclass[pra,twocolumn,showpacs,amsmath,amssymb,superscriptaddress]{revtex4-1}
\usepackage{graphicx}% Include figure files
\usepackage{amsmath}
\usepackage{amssymb}
\usepackage{array}
\usepackage{SIunits}
\usepackage{color}
\usepackage{dcolumn}% Align table columns on decimal point
\def\ket0{$\left|0\right>$}
\def\ket1{$\left|1\right>$}
\def\ca40{$^{40}\mathrm{Ca}^+$}
\def\T2{$\mathrm{T_2$}}
\def\Pr3{$\mathrm{Pr^{3+}}$}
\def\ket#1{$\left|#1\right>$}

\def\eq#1#2{\begin{equation}\label{Eq:#1}#2\end{equation}}
\def\eref#1{(\ref{Eq:#1})}
\def\fref#1{\ref{Fig:#1}}

\begin{document}

\title{A single ion as a shot noise limited magnetic field gradient probe}
%\titlerunning{Magnetic gradient probe}

\author{A.~Walther}\email[]{walthan@uni-mainz.de}\homepage[]{http://www.quantenbit.de}
\author{U.~Poschinger}
\author{F.~Ziesel}
\author{M.~Hettrich}
\author{A.~Wiens}
\author{J.~Welzel}
\author{F.~Schmidt-Kaler}
\affiliation{Institut f\"ur Quantenphysik, Universit\"at Mainz, Staudingerweg 7, 55128 Mainz, Germany}

\date{\today}% It is always \today, today, but any date may be explicitly specified

\begin{abstract}
It is expected that ion trap quantum computing can be made scalable through protocols that make use of transport of ion qubits between sub-regions within the ion trap. In this scenario, any magnetic field inhomogeneity the ion experiences during the transport, may lead to dephasing and loss of fidelity. Here we demonstrate how to measure, and compensate for, magnetic field gradients inside a segmented ion trap, by transporting a single ion over variable distances. We attain a relative magnetic field sensitivity of $\Delta B/B_0 \sim 5 \cdot 10^{-7}$ over a test distance of 140 \micro m, which can be extended to the mm range, still with sub \micro m resolution. A fast experimental sequence is presented, facilitating its use as a magnetic field gradient calibration routine, and it is demonstrated that the main limitation is the quantum shot noise.
\end{abstract}


\maketitle

\section{Introduction}
Over the past decade, trapped ions have emerged as one of the most promising systems to realize quantum computation, with many key ingredients implemented experimentally, such as single and multiple qubit gates~\cite{Blatt2008}. It appears difficult however, to scale single zone traps significantly beyond the order of ten qubits \cite{Monz2010}. One way to store and manipulate such complex systems on a larger scale with a larger number of ions, is to divide the ion trap quantum computer into several regions - each holding only a few ions - combined with transport of the ions between them \cite{KIELPINSKI2002}. In order to realize this, it is paramount to make the transport between storage regions and processor work with minimum loss of coherence, which has proven to be a challenge so far. In this paper we investigate the transport of single ion qubits and address one of the leading dephasing mechanisms, arising from magnetic field gradients. Previous transport demonstrations \cite{HOME2009} have avoided magnetic field induced dephasing by adding an extra set of qubit levels, which are less sensitive to field changes during transport. This, however, adds extra complexity to the scheme, and also may not work for all ion species. It has also been shown that adding extra identical transport sequences after a $\pi$-inversion can cancel out unwanted phase shifts \cite{Blakestad2009} even when transporting across junctions, however, such extra phase-changing operations does not appear to be feasible when transporting qubits that are a part of larger entangled states. Another approach is to use two physical ions in an entangled state to form one logical qubit in a subspace which is free of decoherence from the magnetic field variations over the transported distance \cite{KIELPINSKI2001}, at the expense of needing twice as many ions. Even in this scenario however, the difference in magnetic field magnitude between the two paired ions needs to be sufficiently low.

\begin{figure}[ht]
\includegraphics[width=0.85\linewidth]{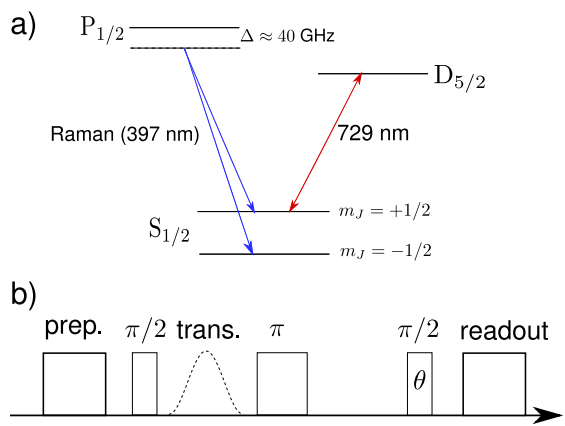}
\caption{ Part a) shows a reduced level diagram of the \ca40 ion with the levels that are relevant for the presented experiments. b) shows the spin echo pulse sequence used to measure the magnetic field gradient. The preparation and the readout each consists of several steps, as described in the text, and the dashed pulse is the electric potential change that causes the transport, while the solid boxes are laser pulses.}
\label{Fig:Scheme}
\end{figure}

Here we present a technique for measuring, and compensating for, the magnetic field gradient inside the trap, using a single ion as a movable probe. We carefully tune the currents of external Helmholtz coils for very low B-field gradients. Stationary two ion crystals also allow for local measurements of gradients~\cite{Chwalla2007}, although the transport method presented here adds further advantages: The attainable accuracy of the measurement is higher, due to the fact that the ion can utilize a much larger distance over which changes to the field are picked up. The gradient is measurable over a much larger spatial extent with a sub-\micro m resolution, and extends to regions outside those accessible by the laser beams and the ion observation, for example into storage regions or even across corners and junctions. 

For general scientific interest we envisage applications of our method where the combination of high spatial resolution and shot noise limited performance is used to map out the magnetic field even of single protons~\cite{Rugar2004} or molecules 
\cite{Knunz2010}. We note that \micro m resolutions have also been reached using Bose-Einstein condensates~\cite{Wildermuth2005}.
%and quantum Hall devices (need reference before we can include it). 

\section{Experimental setup and method}
In our experiment, we employ a segmented micro-structured Paul trap~\cite{SCHULZ2006}, where a combination of static and dynamical electric fields provide confinement for a single \ca40 ion, and where the segmented design of the trap allows for transport~\cite{Huber2010}. The strengths of the trapping potentials lead to harmonic oscillation frequencies of $\omega/2\pi = \left\{1.35, 2.4, 3\right\}$~MHz in axial and both radial directions, respectively. The relevant internal states of the ion are shown in Fig.~\fref{Scheme}a, where Doppler cooling is performed on the $\mathrm{S}_{1/2} \rightarrow \mathrm{P}_{1/2}$ (cycling) transition near 397~nm. The two ground state spin levels, \ket{m_J = \pm 1/2}, are used as qubit states, and operated on by employing stimulated Raman transitions using a separate 397~nm laser, which is detuned by $\Delta \approx 2\pi \cdot 40$ GHz from the cycling transition. In addition, a laser at 729 nm provides state selective shelving during readout, while lasers at 866 and 854 nm prevent trapping in metastable states. Two coils outside of the vacuum chamber in a Helmholtz-like configuration are set up to give a B-field axis at a 45$^\circ$ angle with respect to the trap axis. During the experiment, a magnetic field offset is kept around $\sim0.7$ mT giving a Zeeman splitting of the spin states of about 18 MHz, while gradients are tuned by the application of different currents to the two coils. 

For determination of magnetic gradients we employ the following sequence of operations, depicted in Fig.~\fref{Scheme}b: i) Preparation pulses consisting of Doppler cooling pulses as well as optical pumping pulses initialize the ion to the $\mathrm{m}_J = +1/2$ state. ii) spin echo sequence, with the transport pulse inserted between the first $\pi/2$-pulse and the $\pi$-pulse. iii) readout pulses, which are carried out by an efficient state selective transfer on the 729~nm transition followed by fluorescence from the Doppler laser~\cite{POSCHINGER2009}

Precise knowledge of the transported distance as a function of time comes from a calculation of the potential minimum of the trap for each value of an applied voltage by means of a numerical field solver~\cite{Singer2010a}. In ref.~\cite{Huber2010} we experimentally confirmed that the potentials obtained from the simulations match the actual potentials in our trap with an accuracy on the 1\% level. From the solver one can obtain normalized shape factors, $\tilde{u}_i(x)$, that are fixed by the geometry of the trap and depend on the segment number, $i$, and the distance from that segment, $x$. The total potential can then be obtained by summation of the voltages applied to each segment multiplied by its respective shape factor:
\eq{total_potential}{
	V(t) = V_1\tilde{u}_1(x) + V_2(t)\tilde{u}_2(x).
}
Here, $V_1$ is the fixed potential on the main trapping segment and $V_2(t)$ is a time dependent sine-shaped voltage that is applied to a neighboring segment, such that the ion is moved a certain distance and then moved back ending at it's initial position. $V_2(t)$ is created by a waveform generator and all unused segments are kept at ground. The duration of the transport pulse is T$=400$ \micro s, which is shorter than the coherence time of the qubit state but much longer than the inverse of the axial trap frequency, $(1.35 \mathrm{MHz})^{-1}$. The latter condition means that the ion may adiabatically follow the changes to the voltages, such that the position of the ion can be obtained from the minimum of the total potential at any given time as a function of the applied segment voltages, i.e. $x_{\mathrm{ion}} = x_{\mathrm{ion}}(V_1, V_2(t))$. The total phase shift acquired by the ion is given by the time integral of the frequency shift caused by the B-field gradient during the transport. The frequency shift is proportional to the B-field, which in turn is proportional to the spatial position in a first order approximation. We can thus express the phase shift in terms of the ion position and the magnetic field gradient:
\eq{posint}{
	\phi_{rad} = \frac{g_J \mu_B}{\hbar}\frac{\delta B}{\delta x} \int_0^T{\left[x_{\mathrm{ion}}(t) - x_{\mathrm{ion}}(0)\right]dt},
}
where the Land\'e factor of the $\mathrm{S}_{1/2}$ state is given by $\mathrm{g_J} = 2$. The integral in this equation is thus determined from the displacement of the ion, which is given by evaluating the minimum of the applied potential. It is then clear that the magnetic field gradient, $\delta B/ \delta x$, can be extracted by measuring the acquired phase shift for a particular known displacement-time integral.
\begin{figure}[ht]
\includegraphics[width=8.5cm]{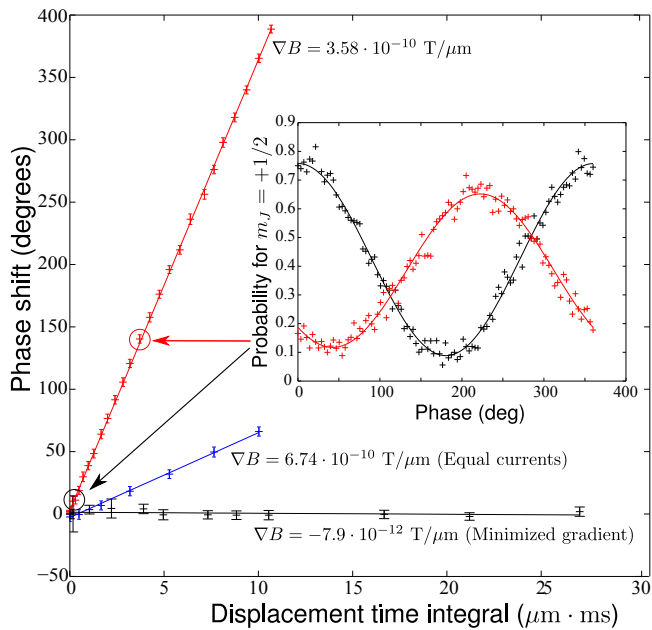}
\caption{(color online) Recorded phase shift as a function of transported distance and time (see text), for three different coil currents. The slope gives the magnetic field gradient, which is listed next to each respective curve. Note how the optimized coil current setting has a gradient that is two orders of magnitude lower than when the coils have the same current. The inset shows the evaluation of a spin echo fringe for one particular gradient and transport setting compared to a non-shifted one. Each point in the inset consists of 200 ion interrogations.}
\label{Fig:all_grads}
\end{figure}

\section{Results}
\subsection{Measurement of magnetic field gradients}
The result of our experiment to measure the magnetic field gradient over a maximum distance of 140 \micro m is plotted in Fig.~\fref{all_grads} for different currents on the Helmholtz coils. Three distinct cases were measured: i) a large gradient corresponding to 3 A in coil 1 and 1 A in coil 2, ii) equal currents, 1.9 A, in each coil and iii) with currents chosen to minimize the gradient, yielding 1.660 A in coil 1 and 2.122 A in coil 2. The particular values in the optimized last case were chosen by extrapolation from the first two cases, with the desire to reach a zero gradient. In all three cases, the offset magnetic field is close to the same value, i.e. around 0.7 mT, but the gradient varies over more than two orders of magnitude. Furthermore, the standard deviation of the phase measurement in Fig.~\fref{all_grads} is about $2^\circ$, and from the near-perfect linear fit, we can conclude that the linear approximation that was made in Eq.~\eref{posint} was well justified. In the figure, each phase shift is evaluated from a cosine fit to the fringe pattern data, resulting from a spin echo sequence where the phase of the final $\pi/2$-pulse is varied between 0 and $2\pi$ in 100 steps, as shown in the inset. The phase is then plotted versus the displacement-time integral (see Eq.~\eref{posint}), i.e. the representation of how long the ion stays in a magnetic field that is different from its starting location, times the transported distance.

\subsection{High precision detection of small gradients}
The main application for this type of magnetic field inhomogeneity measurement is the creation of a low decoherence environment for quantum computing, and we now focus on measurements of the low gradient case with high precision. We also aim at using as short as possible total measurement time, such that the method can be efficiently used as a calibration routine. Shortening the time of the acquisition can be achieved by reducing the number of points that are used in order to determine the acquired phase shift. In principle, for small phase shifts, it is enough to measure the spin up probability at a concluding angle of $\theta=\pi/2$ in the cosine curve, where the slope is the steepest, in order to obtain an estimation for the phase shift. In reality however, various mechanisms can cause offsets and loss of contrast of the fringe, which leads to the $\pi/2$ point getting shifted for other reasons than acquired phase. To accommodate for this, we additionally measure the baseline and the contrast of the fringe pattern by also probing the points $\theta = 0$ and $\theta = \pi$ respectively. We can now form a normalized signal, $S$, which is independent of contrast and offset:
\eq{signal}{S(\phi) = \frac{p(\pi/2,\phi) - p(\pi,\phi)}{p(0,\phi) - p(\pi,\phi)}.
}
With a 5 ms detection time, one measurement of the signal $S(\phi)$, comprised of the three measurement points at $\phi= \{0, 0.5, 1\} \pi$, thus takes 15 ms. In Eq.~\eref{signal}, $p(\theta,\phi)$ denotes the probability of measuring the ion in spin up as a function of the angle of the concluding $\pi/2$-pulse, $\theta$, and the phase shift induced by the magnetic field gradient, $\phi$. In general, $p(\theta,\phi)$ is described by a sinusoidal function with a scaling factor and an offset, i.e. $p(\theta,\phi) = A_1 \cos(\theta + \phi) + A_2$. Putting this expression into Eq.~\eref{signal} however, it becomes clear that the parameter $S$ is independent of both the scaling factor and offset and we find that the phase shift can be obtained as
\eq{phi_s}{
	\phi = \arctan(1-2S).
}
In order to investigate the precision obtained from the above construction, we measured a repeated acquisition of these three points with and without transport, for small changes to the current of the coils. An example of this is shown in Fig.~\fref{Allan}a, where each point is an average of all preceding points, and each error bar is the deviation of the last (averaged) point with respect to the previous points. If the measurement is quantum shot noise limited, this error should converge towards a correct estimation of the phase shift with a rate proportional to $\sqrt{N}$, where $N$ is the number of repetitions of the experiment.
\begin{figure}[ht]
        \includegraphics[width=8.5cm]{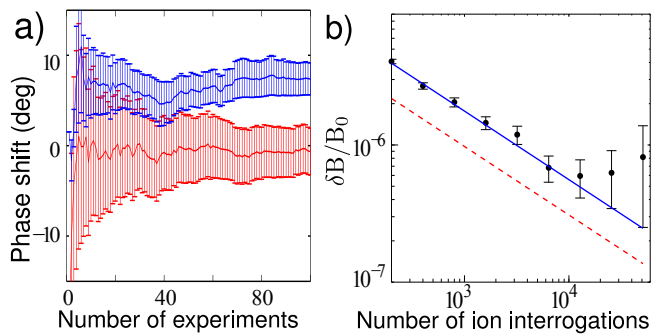}
    \caption{(color online) Part a) shows an example measurement of a small magnetic field difference of 7 nT over 19 \micro m (blue curve) in comparison to a minimized gradient (red curve), where each of the 100 experimental points consists of 200 ion interrogations. In b), an Allan type standard deviation over a larger data set (black dots) is plotted in comparison with the theoretical expectations using the values of the contrast going from 0.12 to 0.75 as realistic values (blue solid line) and using a perfect contrast (dashed red line), for the standard deviation coming from the quantum projection noise limit (see text). The y-axis scale represents the smallest magnetic field difference we can measure on top of the background of B$_0$ = 0.7 mT.}
    \label{Fig:Allan}
\end{figure}

\subsection{Allan variance and the quantum shot noise limit}
While Fig.~\fref{Allan}a demonstrates how the phase shift with and without transport directly can be used to resolve small differences in the magnetic field, an Allan type evaluation of the standard deviation gives a better estimate for the true rate of convergence. This is displayed in part b) of Fig.~\fref{Allan}. In order to show that the main limitation of the experiment comes from quantum shot noise, we derive an expression for the quantum projection noise, caused by the projection on the spin states. This binary choice indicates that the uncertainty for each individual ion interrogation is given by the binomial distribution:
\eq{binom_std}{
	\sigma_{\mathrm{p}}(\phi) = \sqrt{p(\phi) \cdot \left(1-p(\phi)\right)/N},
}
where $p$ is the probability of measuring spin up, and $N$ is the total number of individual interrogations. As each data point consists of three actual measurements, the contributions to the error from all three projections must be taken into account. Gaussian error propagation then leads to an expression for the total error of the phase estimation:
\eq{phase_err}{
	\sigma_{\phi} = \sqrt{\sum_{\phi = 0, \pi/2, \pi} \left(\frac{\delta \phi}{\delta p(\phi)} \sigma_{\mathrm{p}}(\phi)\right)^2},
}
where the partial derivatives can be obtained from Eqs. \eref{signal} and \eref{phi_s}. The binomial standard deviations from Eq.~\eref{binom_std} are used as errors together with values for $p(\phi)$ corresponding to having the three points come from a signal with realistic contrast going from 0.12 to 0.75 ($A_1=0.31$ and $A_2=0.44$), like the one seen in the inset to Fig.~\fref{all_grads}. We then obtain a limit of $\Delta \phi = 1.81 \, \, \mathrm{rad}/\sqrt{N}$, which is plotted in Fig.~\fref{Allan}b, converted into a corresponding relative magnetic field difference. In addition, the figure also displays the minimal obtainable error that could be reached if the spin echo contrast would be perfectly 1. We find that the experimental data follows the quantum projection noise curve, which takes the limited contrast into account. From the above derivation it is also clear that the obtained accuracy is not limited by the time it takes to perform the acquisition, but only by the number of interrogations, which is substantiated by the close proximity to the theoretical limit. This means that technical improvements, such as more efficient camera read out, can be used to reduce the time it takes for a full acquisition. As a reference, the plot in Fig.~\fref{Allan}a contains in total $20000$ ion interrogations (for each of the three points), and with 5 ms detection time per interrogation it took only in the order of minutes to acquire. After about $10^4$ interrogations, we reach a relative magnetic field sensitivity of $\delta B/B \sim 5 \cdot 10^{-7}$. At this point the measured Allan deviation appears to deviate from the theoretically expected one. This behavior is attributed to long term drifts of the experimental setup arising from e.g. temperature variations in the magnetic field coils. The observed contrast loss can be explained by technical noise from the waveform generator that applies the transport pulse voltage, which leads to a small amount of motional heating. This problem can be solved by a better suited multi-channel voltage source that can address all of the trap control segments. Time dependent magnetic field fluctuations faster than the spin echo time is further reducing the contrast, and such noise may be reduced by a passive shielding box of $\mathrm{\mu}$-metal around the trap setup. It can also be mentioned that it is possible to improve the sensitivity of the method by utilizing two or more ions in a spin-squeezed state, which can reduce the noise of the phase measurement even below the shot noise limit~\cite{Meyer2001}.

Our value of the relative magnetic field sensitivity is comparable to other methods used for measuring magnetic fields, such as squids~\cite{Faley2004} or atomic vapors~\cite{Groeger2006,Kominis2003}. Though it may be noted that the other methods target different applications, and while those can reach higher absolute sensitivity, our method relies only on a single ion, and can thus yield a very high spatial resolution, in the sub-\micro m range. Another potential application is the detection of the magnetic field from a single spin. Although it is currently difficult to move the ion close enough ($\sim100$ nm) to the spin for adequate absolute sensitivity, it is interesting to note that our method has a similar relative sensitivity as magnetic resonance force microscopy~\cite{Rugar2004}, where single spin resolution has been reached.

\section{Conclusion}
In conclusion, we have demonstrated a technique to use transport of a single ion to detect magnetic field gradients with high resolution, both spatially and magnitude wise. Using the transport of the ion gives the natural advantage that magnetic gradients in the entire trap, even regions inaccessible to the necessary lasers, can be minimized, which is expected to be very useful for ion traps used for scalable quantum computing. For a single transported ion, the minimum detectable phase shift suggests a transported coherence time would be more than ten ms, which is already long enough that the gradient is no longer the main limitation. The expected coherence time is even significantly longer if we consider a logical qubit in a decoherence free subspace~\cite{KIELPINSKI2001}. This entangled two-ion logical state is insensitive to magnetic field inhomogeneities, although a difference in the magnetic field between the two ions will still cause dephasing. Using the minimized gradient demonstrated here and assuming a reasonable ion separation of 4 \micro m, we find a coherence time above 1 s. This time is long compared to the time of qubit gates as well as the expected transport durations, however, it is not so long that one does not realize the necessity of performing a compensation routine, like the one suggested here, for minimizing the magnetic field inhomogeneity.

\section*{Acknowledgements}
The authors would like to thank Kilian Singer for helpful discussions. We acknowledge financial support by the IARPA project, and by the European commission within the IP AQUTE.

%\vspace{0.2cm}

\bibliographystyle{apsrev4-1}
\bibliography{MagGrad_bib}

\end{document}